\documentclass[usenatbib,useAMS,a4]{mn2e}
\usepackage[dv ips]{graphicx}
\usepackage{natbib}
\usepackage{lastpage}
\usepackage{subfigure}

\title{Circumstellar dust as a solution to the red supergiant supernova progenitor problem}

\author[Joseph J. Walmswell, John J. Eldridge]{Joseph J. Walmswell \thanks{E-mail: jjw49@ast.cam.ac.uk}, John J. Eldridge \\
Institute of Astronomy, The Observatories, University of Cambridge, Madingley Road, Cambridge, CB3 0HA\\}

\date{August 2011}
\pagerange{\pageref{firstpage}--\pageref{lastpage}}
\begin{document}

\maketitle
\label{firstpage}

\begin{abstract}
We investigate the red supergiant problem, the apparent dearth of Type
IIP supernova progenitors with masses between 16 and ${30\,\rm M_{\odot}}$. Although red supergiants with masses in this range
have been observed, none have been identified as progenitors in
pre--explosion images. We show that, by failing to take into account the
additional extinction resulting from the dust produced in the red
supergiant winds, the luminosity of the most massive red supergiants
at the end of their lives is underestimated. We re--estimate the
initial masses of all Type IIP progenitors for which observations
exist and analyse the resulting population. We find that the most
likely maximum mass for a Type IIP progenitor is
$21^{+2}_{-1}\,\rm M_{\odot}$. This is in closer agreement with the limit
predicted from single star evolution models.
\end{abstract}


\section{Introduction}

Nothing in the life of a massive star becomes it like the leaving
it. A supernova (SN) is one of the most impressive spectacles that the
Universe can afford an astronomer. However, there is some uncertainty
as to the range of stars that undergo this most spectacular
demise. Observations indicate that stars evolve into red supergiants
if their initial masses are between about 12 and ${30\,\rm M_\odot}$
\citep{Levesque2006}. Stars that are less massive undergo second
dredge--up and end their lives as asymptotic giant branch (AGB) stars
\citep{2007MNRAS.376L..52E}. Mass--loss rates increase with mass and
stars that are more massive suffer enough mass--loss to remove their
hydrogen envelopes before they die. They become Wolf--Rayet stars, naked
helium stars with thick winds \citep{Crowther}. These explode as
hydrogen--free Type Ib/c SNe, rather than the more common Type II SNe
that are believed to result from the death of red supergiants
\citep{Filippenko,Smartt2009}. Type II SNe are in turn divided into
Types IIP, IIL, IIn and IIb. Types IIP and IIL are identified by their
SN light curves. The former have a plateau and the latter show only
a linear decline. The plateau is the result of the photosphere
maintaining a constant radius, moving inward in mass as the ejecta
expands \citep{Filippenko}. This in turn is due to the ionised
hydrogen recombining. Type IIb SNe have weak hydrogen lines and light
curves similar to hydrogen--free Type Ib SNe, implying that they
contain only a small percentage of hydrogen. There seems to be a
sequence from Type IIP to Type IIL to Type IIb SNe, driven by
increased mass-loss and a consequently reduced mass of hydrogen in the ejecta. Type IIn SNe
are distinguished by narrow line hydrogen emission, the result of
shock interaction with circumstellar material
\citep{2011MNRAS.412.1522S}. The mass limits have some dependence on
metallicity because metal--rich stars have higher mass--loss rates and thus
the minimum initial mass of Wolf--Rayet stars is approximately ${25\,\rm
  M_\odot}$ at Solar metallicity \citep{2004MNRAS.353...87E}. We
therefore expect Type II progenitors up to this limit.

The red supergiant problem was first reported by \citet{Smartt2009}.
They compared 20 Type IIP SN progenitor detections and non--detections
with stellar evolution models to determine the minimum and maximum
initial mass limits for the progenitors of these SNe. They found that
the minimum mass required for stars to explode as Type IIP SNe was
${\rm 8.5_{-1.5}^{+1} \, M_\odot}$, which is consistent with the
observed upper limit for white dwarf formation
\citep{Weidemann,2009ApJ...693..355W}. More surprisingly, they found
an upper limit of ${\rm 16_{-1.5}^{+1.5} \, M_\odot}$, a 95 per cent
confidence upper limit of ${\rm 21 \, M_\odot}$. In essence, the red
supergiant problem is that this estimate is well below the maximum
mass estimated for red supergiants. There therefore appears to be an
absence of higher--mass red supergiant SN progenitors, leaving the fate
of stars with masses between 16 and 25-${30\,\rm M_\odot}$ uncertain.

It should be mentioned that the alternative to deducing masses from
progenitor models is to model the supernova directly. This has the
advantage of not requiring pre--explosion images, although it does
depend on detailed follow--up observations of the supernova
itself. Utrobin and Chugai have used a one--dimensional hydrodynamic
code \citep{2004AstL...30..293U} to model the SNe 2005cs
\citep{2008A&A...491..507U} and 2004et \citep{2009A&A...506..829U}.
Both have detected progenitors and in this paper we deduce initial
masses of $9^{+1}_{-4}$ and ${12^{+1}_{-1}\,\rm M_\odot}$
respectively. This compares with 18 and ${28\,\rm M_\odot}$ for the
hydrodynamic masses. If these masses are correct, the red supergiant
problem ceases to be. However, three earlier independent attempts to
model the progenitor of 2005cs gave masses of ${\rm 9^{+3}_{-2} \,
  M_\odot}$ \citep{2005MNRAS.364L..33M}, ${\rm 10^{+3}_{-3} \,
  M_\odot}$ \citep{2006ApJ...641.1060L} and between 6 and ${8\,\rm
  M_\odot}$ \citep{2007MNRAS.376L..52E}, a discrepancy noted by the
authors. There are other approaches; \citet{2011MNRAS.410.1739D} have
constructed models that include the nebula phase of the supernovae,
which is more amenable to detailed simulation. These SNe modelling
approaches have great potential but wait on more sophisticated models.

Another approach is to consider the ratios of the different types of
SNe. One then integrates the initial mass function (IMF) to find the
limits that provide the desired number of
stars. \citet{2011MNRAS.412.1522S} considered the fractions of core
collapse SNe from the Lick Observatory Supernova Search (LOSS) and
observed that the proportion of Type Ib/c SNe was much too high for
all of them to be the result of single star evolution.  In addition,
with the standard Salpeter IMF, the observed fraction of Type IIP SNe
was such that it could be produced by single stars with initial masses
in the range ${8.5-–13.7 \rm M_\odot}$. These facts imply that binary
interaction allows the production of hydrogen--free progenitors at
lower masses than would otherwise be the case.
\citet{2011MNRAS.412.1522S} suggest that if binaries are included the
upper limit for red supergiants would be in the range of 18 to 24
${\rm M_{\odot}}$.

However, this study found a lower fraction of Type IIP SNe than
previous work. The core collapse SNe broke down as 48 per cent Type
IIP and 22 per cent Type Ib/c, compared with 59 and 29 respectively
for the survey of \citet{2009ARA&A..47...63S}. The reason for this is
not clear. \citet{2009ARA&A..47...63S} considered a 28 Mpc
volume--limited survey using all detected SNe within that volume,
whereas \citet{2011MNRAS.412.1522S} used 60 Mpc and only those SNe
detected in LOSS. Type IIP SNe tend to be dimmer than other types so
those at large distances may have been missed. While these selection
effects were accounted for, such adjustments are, by their nature,
very uncertain. Another reason for the difference could be that
\citet{2011MNRAS.412.1522S} had more complete spectroscopic and light
curve follow-ups and so had greater accuracy in their classifications.

It is possible to obtain the required SNe fractions with an
appropriately chosen stellar population. \citet{2011MNRAS.tmp..692E}
showed that a population composed of a mixture of binary and single
stars could explain the rates of
\citet{Smartt2009}. \citet{2011MNRAS.412.1522S} used their own binary
population models to explain their results. The uncertainity in the
rates means that it is hard to use these fractions to tightly
constrain the progenitor mass range, particularly when the shape of
the IMF means that small changes in the lower limit result in large
changes in the population fractions. In contrast large changes in the
upper limit result in small changes in the population
fractions. Hopefully future surveys will resolve the discrepancy.

If we accept the existence of the red supergiant problem then we must
consider a number of possible explanations. First, that these massive
red supergiants form black holes with faint or non--existent SNe
\citep[e.g.][]{2003ApJ...591..288H}. Secondly, that their envelopes
are unstable and eject a large amount of mass pre--SN
\citep[e.g.][]{2010ApJ...717L..62Y}. Thirdly, that they explode as a
different type of SN \citep[e.g.][]{2006A&A...460L...5K}. We suggest a
fourth explanation, that the mass estimates may be systematically
inaccurate at the high--mass end. Mass estimates are based on
mass--luminosity relations, so extra intrinsic extinction close to the
red supergiant progenitors would give reduced luminosities and hence
lower predicted masses. \citet{Smartt2009} were careful to provide
extinction estimates when possible from measurements of nearby
supergiants and from the supernova itself. These could be
underestimates. If red supergiants produce extra dust local to the
star, it would be destroyed in the supernova explosion. This is very
plausible. It is known that red supergiants form dust in their winds
\citep{2005ApJ...634.1286M}. Furthermore, IR interferometry has shown
that this dust can be found very close to the star itself
\citep{DanchiBester}. \citet{2011MNRAS.412.1522S} also
  suggest dust as a solution to the red supergiant problem.

In this work, we first describe the theoretical models from which we
derived mass--magnitude relations. We then show how these were used to
deduce masses from a population of Type IIP SN progenitors. Finally,
we deduce the most probable upper and lower mass limits and present our
conclusions.

\section{Simulating red supergiants}

We begin by considering the Cambridge ${\sc \rm STARS}$ code, the source
of our stellar models. These models were processed to generate
observed colours with the BaSeL V3.1 model atmosphere grid
\citep{Westera} and the relevant broad--band filter functions. We used
the dust production rate observed by \citet{2005ApJ...634.1286M} to
estimate the amount of circumstellar extinction that would then
manifest.  This allowed the calculation of mass--magnitude relations
both with and without the inclusion of circumstellar dust. We also
consider the nature of dust production, including non--spherical
behaviour.

\subsection{The Cambridge ${\sc \rm STARS}$ code}

The Cambridge ${\sc \rm STARS}$ code was originally developed by Peter Eggleton in
the 1960s \citep{Eggleton}. It uses a non--Lagrangian mesh, where the
mesh function ensures that the points are distributed so that no
quantity of physical interest is allowed to vary by a large amount in
the intervals. The code has been gradually improved and updated and
the work in this paper is based on the code described by
\citet{Stancliffe} and those referenced by them.

Convection is included in the code by the mixing length theory of
\citet{BohmVitense}, with a solar--calibrated mixing length parameter
of $\alpha=2.0$. Convective overshooting is obtained with the method
of \citet{Schroder}, with an overshooting parameter of
${\rm \delta_{OV}=0.12}$. This method involves the addition of a ${\rm \delta}$
term to the adiabatic gradient, allowing mixing to occur in regions
that are weakly stable by the Schwarzchild criterion. The code follows the chemical evolution of ${\rm^{1}H}$, ${\rm ^{3}He}$,
${\rm ^{4}He}$, ${\rm ^{12}C}$, ${\rm ^{14}N}$, ${\rm ^{16}O}$ and ${\rm ^{20}Ne}$, together with structural variables. 

We use the mass--loss scheme described by
\citet{2004MNRAS.353...87E}. For main--sequence OB stars the mass--loss
rates are calculated according to \citet{2001A&A...369..574V} and for
all other stars we use the rates of \citet{1988A&AS...72..259D}, where
the metallicity scaling goes as ${\rm (z/z_{\odot})^{0.5}}$. This
theory is older but the the rates have been recently tested for red
supergiants and been shown to be still the best rates available for
them \citep{2011A&A...526A.156M}.

We have created a library of evolution models, with values of $Z$, the
metallicity fraction by mass, equal to 0.02, 0.01, 0.008 and
0.006. These cover the metallicity range of the IIP progenitors given
by \citet{Smartt2009}. The fractions of hydrogen and helium were
determined on the assumption of constant helium enrichment from the primordial
condition of $X$=0.75, $Y$=0.25 and calibrating to a Solar composition of
$X$=0.70, $Y$=0.28 and $Z$=0.02, i.e. that $Y=0.25+1.5Z$. We evolve our
models through to core neon burning, a few years before core collapse.

\subsection{The colour--magnitude diagrams}

The Cambridge ${\sc \rm STARS}$ code outputs the physical parameters for
a stellar model at each time--step. These include the bolometric
luminosity and the surface temperature. Because the stellar observations
are in the form of colours and magnitudes, either the luminosity and the surface
temperature must be estimated from the observations or a method of
calculating colours and magnitudes from the models must be developed. We
choose the latter course because this requires fewer
assumptions. \citet{Smartt2009} used the opposite approach and, for
that reason, the masses they deduced sometimes differ by a small
amount. We use the method described by \citet{EldridgeStanway2009} to
process the code output and calculate magnitudes to compare to those
observed. The BaSel v3.1 grid of model atmospheres is arranged over
surface temperature and effective gravity. Using the values of these
variables from the code, we obtained appropriate templates for the
SEDs for each model at each time--step. We then applied the filter
functions to extract the magnitudes in the various
bands. This allowed us to produce colour--magnitude diagrams from
the evolution tracks.

\begin{figure}
\centering    
\includegraphics[width=0.45\textwidth]{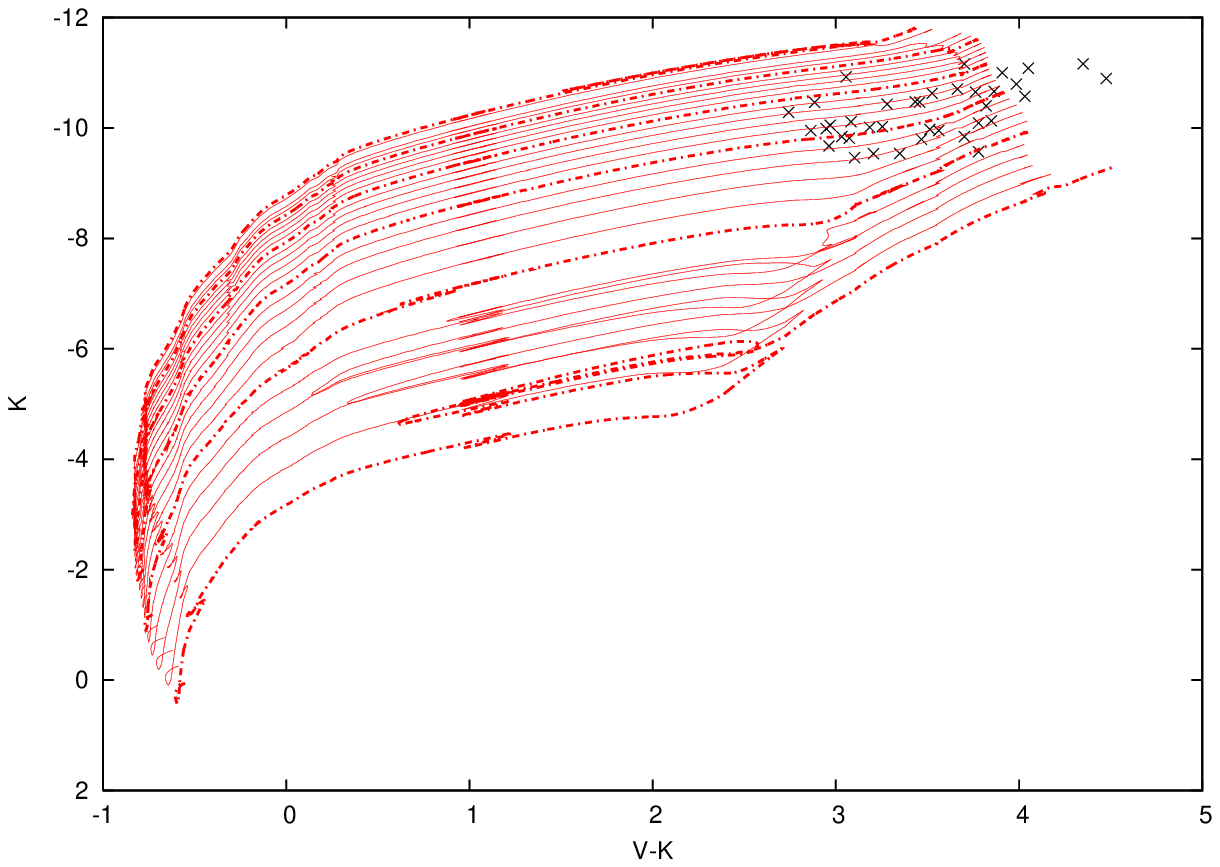}
\includegraphics[width=0.45\textwidth]{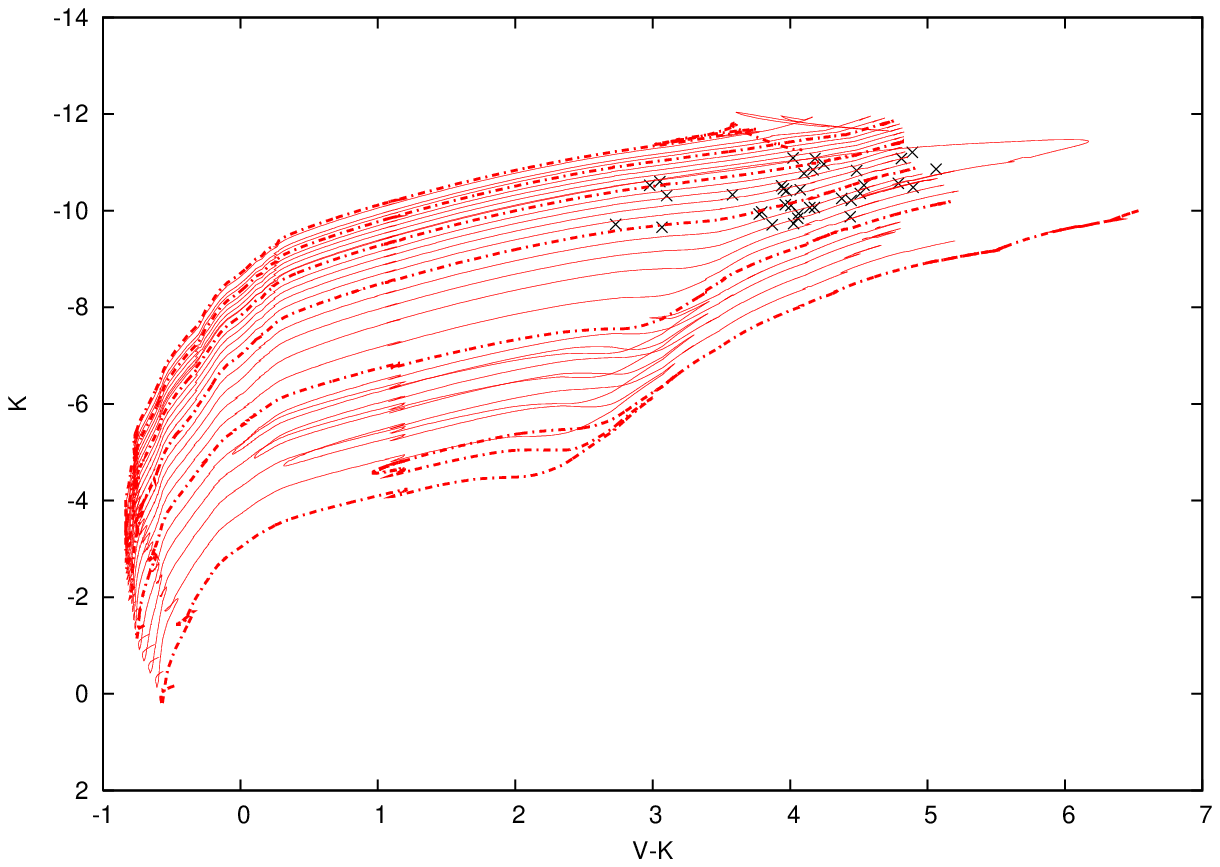}
\includegraphics[width=0.45\textwidth]{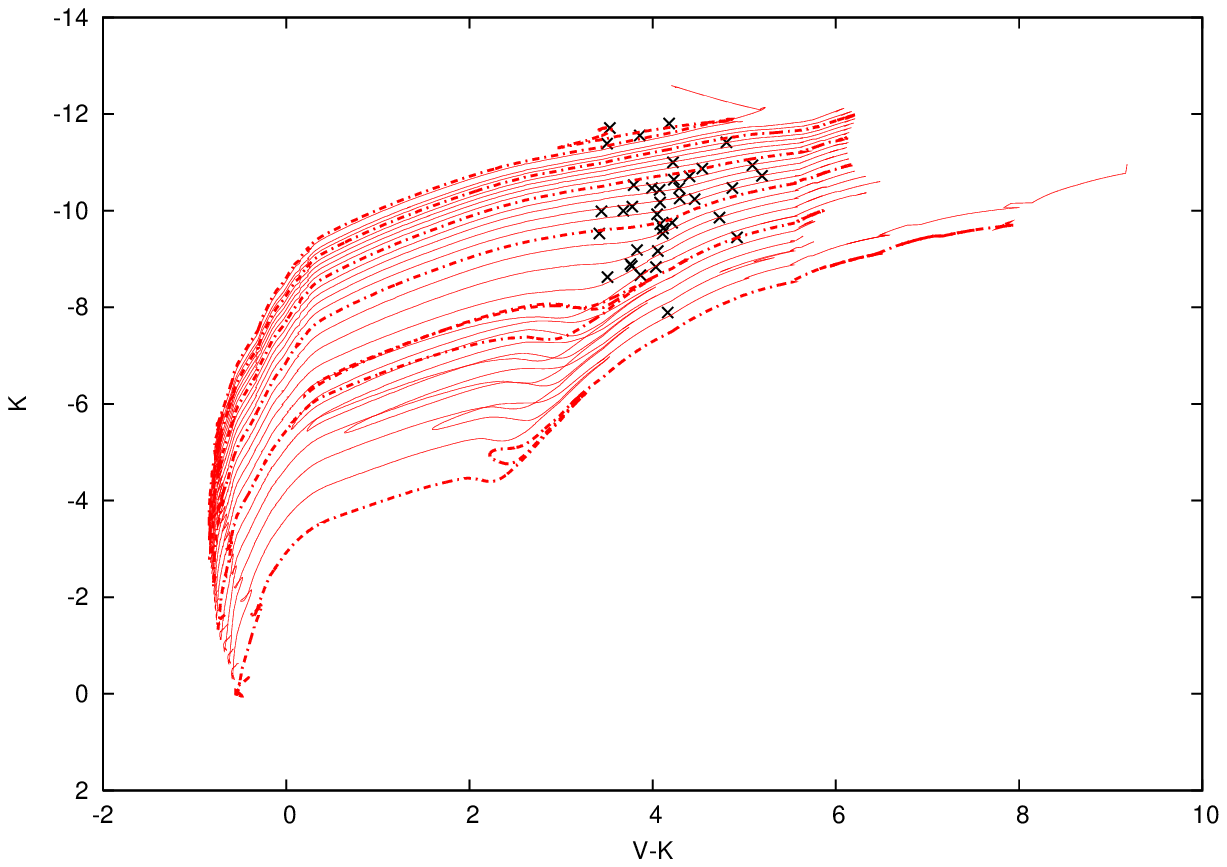}
\caption{Evolution tracks for every integer mass between 5 and
  ${30\,\rm M_\odot}$, with the multiples of 5 indicated by dashed
  lines. The crosses are the observed red supergiants. The SMC stars
  are at the top and the models use $Z=0.004$. The LMC stars are in
  the middle and the models use $Z=0.01$. The Milky Way stars are at
  the bottom and the models use $Z=0.02$.}
\label{fig:redsupergiants}
\end{figure}

To check the validity of our synthetic colours we compared our models
to red supergiants observed in the Magellanic Clouds by
\citet{Levesque2006} and in the Milky Way by
\citet{2005ApJ...628..973L}. These observations included estimates for
the total extinction based on spectrophotometric modelling, so it was
possible to process the data to get the absolute colours and
magnitudes in the absence of extinction. This meant that the models
did not yet have to take into account the effects of circumstellar
extinction. This is shown in Figure~\ref{fig:redsupergiants}. We see
that the models agree well with the observations, with the red
supergiants appearing towards the end of the evolution tracks and
ranging in mass between about 10 and ${30\,\rm M_\odot}$. Some of the
SMC stars are cooler than predicted and may have higher
metallicities. This is not unexpected with a large and heterogeneous
population.

\begin{figure}
\centering \includegraphics[width=0.5\textwidth]{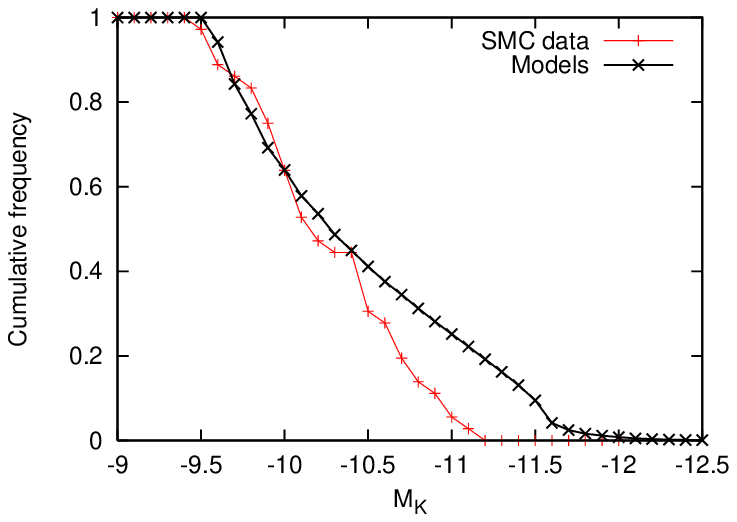}
\centering \includegraphics[width=0.5\textwidth]{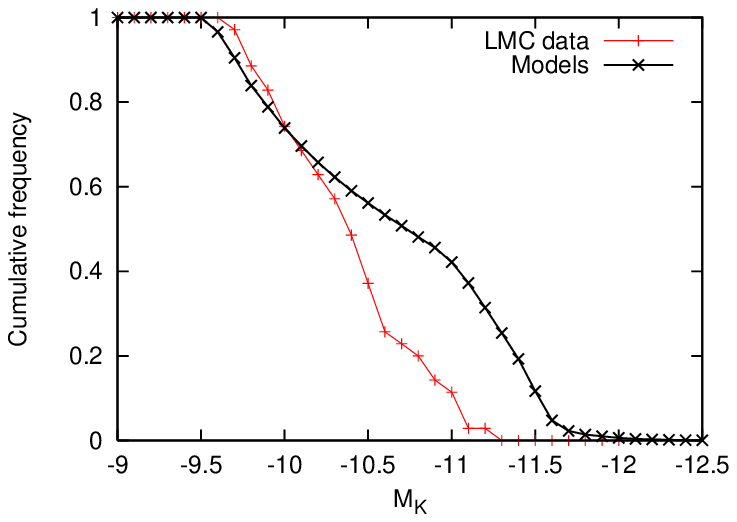}
\centering \includegraphics[width=0.5\textwidth]{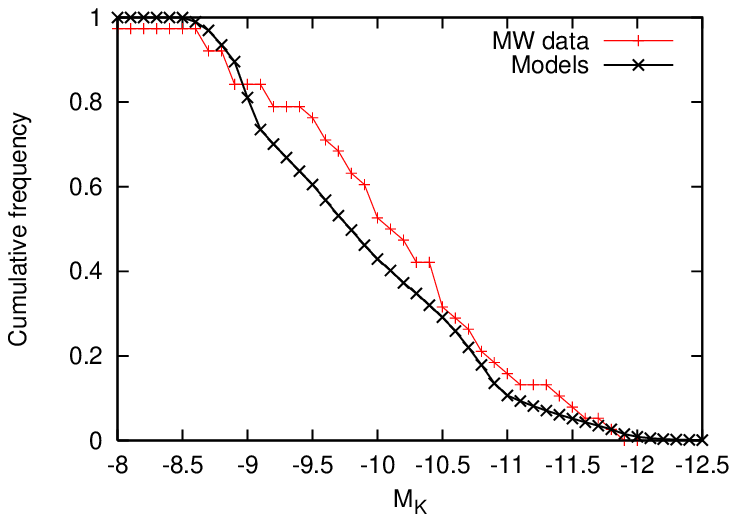}
\caption{The cumulative frequency diagrams, in $K$ band magnitude, for
  red supergiants observed in the SMC, the LMC and the Milky way. They
  are compared with cumulative frequency curves from synthetic
  populations derived from the same models used for
  Figure~\ref{fig:redsupergiants}}
\label{fig:cumul}
\end{figure}

To test the validity of the models further, we took the data and made
cumulative frequency plots in $M_K$. We made similar plots from the
models by first identifying which stars became red supergiants during
their lifetimes. We required that $V-K>2.5$ so as to get the reddest
stars. The lower limit in $K$ was set to be $-9.5$ for the Clouds
models and $-8.5$ for the Galactic models. These limits are reasonable
for red supergiants and were chosen to reflect the distribution of the
observations. We weighted the $K$ magnitudes by the timestep to
reflect a greater probability of observation, and the Salpeter IMF. We
used metallicities of $Z=0.004$ for the SMC, $Z=0.01$ for the LMC and
$Z=0.02$ for the Galactic population. The comparison is shown in
Figure~\ref{fig:cumul}.

The agreement is quite good, although there are fewer very luminous
stars in the LMC and SMC sets than this simple application of our models
predicts. This is probably because all three data sets were not
compiled to reflect a stellar population but as an observationally
convenient group of red supergiants. In addition, because we weight by the
timestep and the IMF only, we are implicitly assuming a constant rate
of star formation, a doubtful supposition. We have only considered
single stars and binary interactions may change these predicted
synthetic frequencies \citep{2008MNRAS.384.1109E} and we may be
underestimating the mass--loss rates of the most luminous red
supergiants \citep{2010ApJ...717L..62Y}.

\subsection{Circumstellar dust}

When the surface temperature of a red supergiant falls below about
5000 K, dust begins to condense out of the stellar wind at a distance
of around ${\rm 5-10 R_{star}} \approx 1000 R_{\odot}$
\citep{Massey2009conf}. It might be expected that the amount of dust
production would correlate with the mass--loss, which in turn
correlates with the luminosity, because this is responsible for the
stellar wind \citep{VanLoon}. \citet{2005ApJ...634.1286M} showed that the dust
production rate correlates with the bolometric luminosity, with a
least squares fit of ${\rm log_{10}(\dot{M_{dust}})=-0.43M_{bol} - 12.0}$,
where the dust production rate has units of ${\rm M_\odot / year}$.

The dynamics of the wind and the dust is complicated and simulations
indicate an absence of spherical symmetry. \citet{Woitke} found that
various instabilities such as Rayleigh--Taylor or Kelvin--Helmholtz lead to the formation of arcs and caps of dust despite the
spherical initial conditions. This means that one would expect
variation in the observed mass--loss and extinction that is entirely due to
the behaviour of the dusty wind along the line of sight. This cannot be
accounted for with the ${\rm \sc STARS}$ code, because in the absence of observations
of the dust, one can only use the relation between the luminosity and
the dust production to estimate an average extinction. The
additional variation due to the lack of spherical symmetry means an
additional source of uncertainty.

The amount of extinction that is due to circumstellar dust depends on the past
history of the star, which has a continuous loss of mass over time. In contrast,
the ${\rm \sc STARS}$ code calculates the properties of the stars at intervals
determined by the time--step.  To account for this, we calculated the
launch velocity of the dust and the dust--production rate for each time
in the output data. This, together with the stellar radii, when
interpolated, gave the distribution of dust with distance from the
star. Following \citet{2005ApJ...634.1286M} we begin by referring to
\citet{Whittet} who showed how the extinction owing a thin shell of
dust can be determined. If one assumes a dust grain density of $s={\rm
  2500\,kg\,m^{-3}}$, applicable to low-density silicates and a
refractive index of $m=1.50$, one can obtain the extinction ${
  A_V}$ in terms of the path length $L$.

\begin{equation}
\rho_d = (3.7\times10^{-8}) s \frac{m^2+2}{m^2-1}\frac{A_V}{L}.
\end{equation}

After the substitutions are made, one obtains the density in terms of the
path length and the extinction only.

\begin{equation}
\rho_d = (3.1\times10^{-4}) \frac{A_V}{L}.
\end{equation}

If a thin shell of dust is used, the thickness of the shell ${\delta R}$ cancels with the path length. The extinction caused
by the shell is then in terms of the dust mass ${M_{d}}$ and the
radius $R$.

\begin{equation}
A_V =\frac{(3.2\times10^3) L M_d}{4 \pi R^2 \delta R}.
\end{equation}
\begin{equation}
A_V =\frac{(3.2\times10^3) M_d}{4 \pi R^2}.
\end{equation}

The dust was modelled as a series of these thin shells, over which the
total extinction was integrated at each timestep. The ${A_V}$ was
then used to calculate the extinction in the other pass--bands by using
the extinction law and associated ratios described in
\citet{Cardelli1989}. These are shown in Table \ref{tab:filter}.

\begin{table}
\caption{Extinction at standard and HST pass--bands relative to the V band, as given by \citet{Cardelli1989}.}
\begin{tabular}{| l |  l | }
  \hline \hline
  Filter & ${\rm A_{\lambda}/A_V}$ \\
  \hline
  U & 1.569 \\
  B & 1.337 \\
  V & 1.000 \\
  R & 0.751 \\
  I & 0.479 \\
  J & 0.282 \\
  H & 0.190 \\
  K & 0.114 \\
  F555W & 0.996 \\
  F606W & 0.885 \\
  F814W & 0.597 \\
  \hline \hline
\end{tabular}
\label{tab:filter}
\end{table}

The density of dust falls off according to an inverse square law as
each shell moves outwards in the stellar wind. Therefore only material
very near to the star has a significant effect. We find about 95 per
cent of the extinction is due to material closer than ${50 \, R_{star}}$ in all the models. This means that the dust is unlikely
to affect the SN because it is rapidly swept up and destroyed in the
explosion. The inverse square law also means that the distance at
which dust first forms in the wind is important. We chose to set this
to ${10\, R_{star}}$, an upper estimate, to ensure that our extinctions are modest underestimates.

\section{The supernova progenitors}

We use the compilation of SN detections and
non--detections of \citet{Smartt2009}. All progenitor information can be found in that paper
and the references therein. We supplement this with SN 2009md \citep{2009md}.
 
The metallicities are based on neighbouring O/H number ratios, where
${\rm [O/H] = log_{10}(O/H)} + 12$. For ${\rm [O/H]>8.7}$, the $Z=0.02$
models were used. Similarly, for ${\rm 8.5<[O/H]\leq8.7}$, $Z=0.01$, for
${\rm 8.4<[O/H]\leq8.5}$, $Z=0.008$ and for ${\rm 8.2<[O/H]\leq8.4}$,
$Z=0.006$. This is a more precise division by metallicity than was
possible for \citet{Smartt2009}. Errors are given, when known, and the
errors for the absolute magnitude were obtained by combining the other
errors in quadrature.

We assigned masses to the progenitors by comparing the absolute
magnitudes, together with their error bars, with a plot of red supergiant final
luminosities against initial mass. We decided to plot the minimum and
maximum magnitudes in the lifetime of our theoretical red supergiants,
the lifetime being between the end of core helium burning and the
termination of the model. This gave a range in luminosity over which a
star of a given initial mass might explode. Masses were first obtained from the original models and again after processing to include
the effects of circumstellar dust. The error in the magnitudes gave
the error in the masses.

\begin{figure*}
\centering
\includegraphics[width=0.45\textwidth]{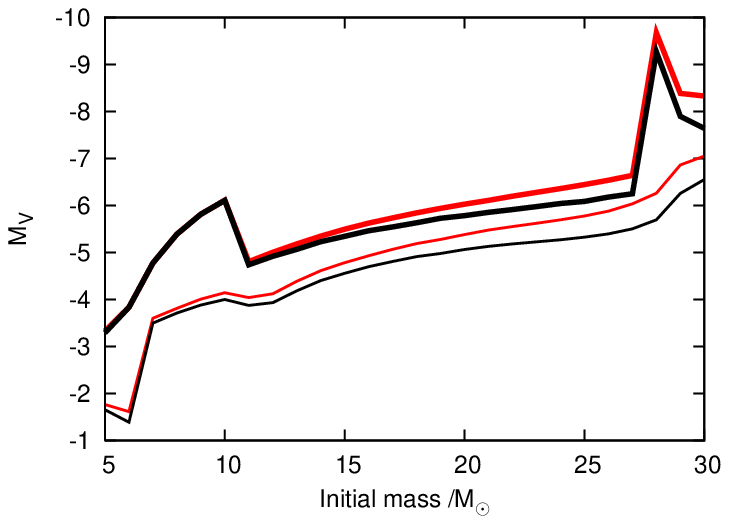}
\includegraphics[width=0.45\textwidth]{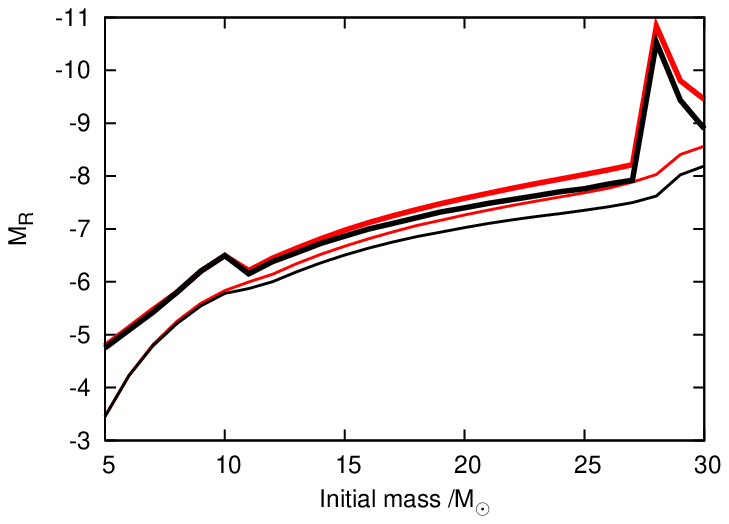}
\includegraphics[width=0.45\textwidth]{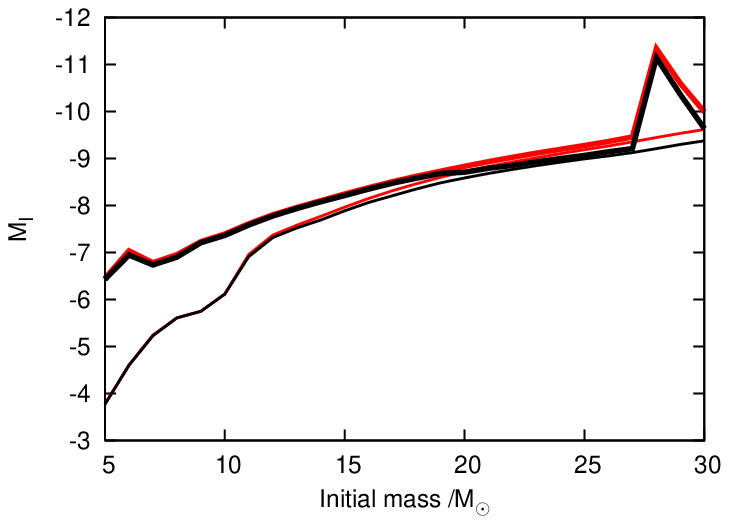}
\includegraphics[width=0.45\textwidth]{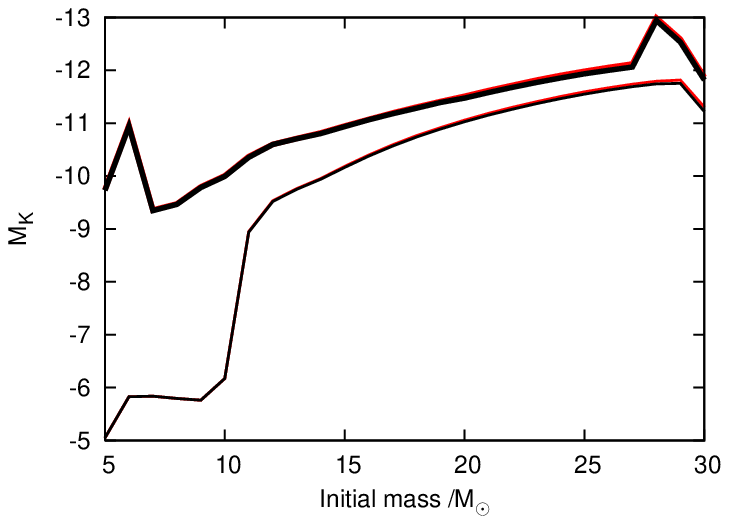}
\caption{The final magnitudes in $V$, $R$, $I$, and $K
$  from our Z=0.02 stellar models. Red indicates the dust free models and
  black the models with dust. The thinner lines are for the minimum magnitudes and the thicker are for the maximum magnitudes.}
\label{fig:MLrelation1}
\end{figure*}

Figure~\ref{fig:MLrelation1} shows how the predicted magnitudes for
the $Z=0.02$ models varies with initial mass in the $V$, $R$, $I$, and $K$
bands. As expected, dust is much more of a problem at shorter
wavelengths, when extinction is more severe, and at higher masses,
when mass--loss is greater.

However, although extinction is less important at longer wavelengths,
changes in stellar luminosity have more of an effect. When a red
supergiant becomes more luminous, most of this increase in output is at
longer wavelengths. One can see from Figure~\ref{fig:MLrelation1} that
the difference between the minimum and maximum magnitudes for a given
mass is highest in the $K$ band. This is particularly acute for the
lower masses because they end core helium burning as blue supergiants,
setting a very low minimum magnitude in the infrared. 

The lower--mass models also have higher maximum magnitudes than the
extrapolation of the behaviour of the more massive stars might
predict. These stars undergo second dredge--up, becoming more luminous
asymptotic giant branch (AGB) stars \citep{2007MNRAS.376L..52E}. In
both cases, increasing the magnitude range also increases the
uncertainty in the progenitor mass--luminosity relation. Finally the
sharp increase in the maximum $V$ magnitude at the high mass end indicates incipient
Wolf--Rayet star formation.

\subsection{The non--detections}

For the non--detections, no progenitor was confidently identified in
the pre--images. This still provides useful information because it sets a
limit on the magnitude of the progenitor. If the star were
brighter, it would have been detected. If the errors in the magnitudes
are approximately normal, there is an 84 per cent probability that the
magnitude of the non--detected progenitor is less than the upper error
bar. This is a sufficiently high confidence level that we took the upper mass
limit to be the lowest mass with a magnitude range entirely
brighter than this upper error bar.

Some of the SNe had non--detections in several pass--bands. The
magnitude limits quoted in Table \ref{tab:nondetection} are those
which gave the lowest upper mass limit. The other pass--bands merely
set a higher upper mass limit which added no additional
information. If we compare the mass limits deduced from the dustless
models with those with dust we can see that the difference
between the two increases with mass. Most notably that for SN 2003ie changes from
22 to ${25 \, \rm /M_\odot}$.

\begin{table*}
\caption{The observed parameters and estimated upper mass limits for Type IIP
  supernova progenitors that were not detected in pre--explosion images. We
  include the masses without considering extinction due to intrinsic dust,
  ${\rm M_{dustless}}$, and the masses taking intrinsic dust into account
    ${\rm M_{dust}}$. Note that we consider 2004A to be a non--detection. The observation of a progenitor was doubtful.}
\begin{tabular}{| l | l | l | l | l | l | l | l | l |}
  \hline \hline
  SN & Metallicity & Distance & Apparent  & Absolute  & ${\rm M_{dustless}}$  & ${\rm M_{dust}}$ \\
     &        /dex & /Mpc     & magnitude & magnitude & ${\rm /M_\odot}$ & ${\rm /M_\odot}$ \\
  \hline
  1999an & 8.3 & 18.5$\pm$1.5  & ${\rm m_{F606W}>24.7\pm0.2}$ & ${\rm M_{F606W}>-7\pm0.3}$ & 18 & 21\\
  1999br & 8.4 & 14.1$\pm$2.6  & ${\rm m_{F606W}>24.91}$ & ${\rm M_{F606W}>-5.89\pm0.4}$ & 11 & 12\\
  1999em & 8.6 & 11.7$\pm$1.0 & ${\rm m_I>22.0}$ & ${\rm M_{I}>-8.5\pm0.2}$ & 18 & 19\\
  1999gi & 8.6 & 10.0$\pm$0.8 &  ${\rm m_{F606W}>24.9\pm0.2}$ & ${\rm M_{F606W}>-5.7\pm0.3}$ & 12 & 13\\
  2001du & 8.5 & 18.2$\pm$1.4 &  ${\rm m_{F814W}>24.25}$ & ${\rm M_{F814W}>-7.4\pm0.2}$ & 13 & 13\\
  2003ie & 8.5 & 15.5$\pm$1.2 &  ${\rm m_{R}>22.65}$ & ${\rm M_{R}>-8.3\pm0.2}$ & 22 & 25\\
  2004A &  8.3 & 20.3$\pm$3.4 &  ${\rm m_{F814W}>24.25}$ & ${\rm M_{F814W}>-7.4\pm0.2}$ & 14 & 15\\
  2004dg & 8.5 & 20.0$\pm$2.6 &  ${\rm m_{F814W}>25.0}$ & ${\rm M_{F814W}>-6.9\pm0.3}$ & 11 & 11\\
  2006bc & 8.5 & 14.7$\pm$2.6 &  ${\rm m_{F814W}>24.45}$ & ${\rm M_{F814W}>-6.8\pm0.5}$ & 11 & 12\\
  2006my & 8.7 & 22.3$\pm$2.6 &  ${\rm m_{F814W}>24.8}$ & ${\rm M_{F814W}>-7.0\pm0.2}$ & 11 & 11\\
  2006ov & 8.9 & 12.6$\pm$2.4 &  ${\rm m_{F814W}>24.2}$ & ${\rm M_{F814W}>-6.3\pm0.4}$ & 10 & 11\\
  2007aa & 8.4 & 20.5$\pm$2.6 &  ${\rm m_{F814W}>24.44}$ & ${\rm M_{F814W}>-7.2\pm0.3}$ & 12 & 13\\
  \hline \hline
\end{tabular}
\label{tab:nondetection}
\end{table*}

\subsection{The detected progenitors}

The actual detections are fewer in number than the non--detections. For progenitors observed in
multiple bands, a mass for each band was calculated and these were
averaged. For SN 2009bk \citet{2008ApJ...688L..91M} used the
progenitor SED to deduce a total extinction of ${A_V=1.0\pm0.5}$
and the absolute magnitudes take this into account. Because this includes any
extinction from circumstellar dust, there was no need to use the dusty
models and the predicted mass is the same in both cases.

\subsection{Other deduced masses}

Table \ref{tab:baddetection} lists the properties of SNe 2004am and 2004dj, both of
which are Types IIP but lack detected progenitors. \citet{Smartt2009}
describe how population synthesis codes were used to deduce progenitor
masses from their parent clusters. These SNe were part of the survey
and are included for completeness.

\begin{table*}
\caption{The observed parameters and estimated masses for Type IIP
  supernovae that were detected in pre--explosion images. We
  include the masses without considering extinction by intrinsic dust,
  ${\rm M_{dustless}}$, and the masses taking intrinsic dust into account,
    ${\rm M_{dust}}$.}
\begin{tabular}{| l | l | l | l | l | l | l | l | l |}
  \hline \hline
  Supernova & Metallicity & Distance & Apparent  & Absolute  & ${\rm M_{dustless}}$ & ${\rm M_{dust}}$ \\
            & /dex         &    /Mpc  & magnitude & magnitude & ${\rm /M_\odot}$ & ${\rm /M_\odot}$ \\
  \hline
  1999ev & 8.5 & 15.14$\pm$2.6  & ${\rm m_{F555W}=24.64\pm0.17}$ & ${\rm M_{F555W}=-6.7\pm0.4}$ & $18^{+3}_{-3}$ & $20^{+6}_{-4}$\\
  2003gd & 8.4 & 9.3$\pm$1.8  & ${\rm m_{V}=25.8\pm0.15}$  & ${\rm M_{V}=-4.47\pm0.5}$  & $8^{+2}_{-1}$ & $8^{+2}_{-2}$\\
         &                        &               &       ${\rm m_{I}=23.13\pm0.13}$ &     ${\rm M_{I}=-6.92\pm0.4}$ & & \\
  2004et & 8.3 & 5.9$\pm$0.4  & ${\rm m_{I}=22.06\pm0.12}$ & ${\rm M_{I}=-7.4\pm0.2}$ & $11^{+1}_{1}$ & $12^{+1}_{-1}$\\
  2005cs & 8.7 & 8.4$\pm$1.0  & ${\rm m_{I}=23.48\pm0.22}$ & ${\rm M_{I}=-6.3\pm0.3}$ & $9^{+1}_{-4}$ & $9^{+1}_{-4}$\\
  2008bk & 8.4 & 3.9$\pm$0.5  & ${\rm m_{I}=22.20\pm0.19}$ & ${\rm M_{I}=-7.2\pm0.4}$ & $12^{+2}_{-4}$ & $12^{+2}_{-4}$\\
    &                        &              &    ${\rm  m_{H}=18.78\pm0.11}$ & ${\rm M_{H}=-9.4\pm0.3}$ & & \\
    &                        &              &    ${\rm m_{K}=18.34\pm0.07}$   & ${\rm M_{K}=-9.7\pm0.3}$ & & \\
    &                        &              &    ${\rm m_{J}=19.50\pm0.06}$  & ${\rm M_{J}=-8.7\pm0.3}$ & & \\
  2009md & 9.0 & 21.3$\pm$2.2 & ${\rm m_{V}=27.32\pm0.15}$ & ${\rm M_{V}=-4.63^{0.3}_{-0.4}}$ & $8^{+4}_{-2}$ & $8^{+5}_{-2}$\\
    &                        &              &   ${\rm m_{I}=24.89\pm0.08}$ &  ${\rm M_{I}=-6.92^{+0.4}_{-0.3}}$ & & \\
  \hline \hline
\end{tabular}
\label{tab:detection}
\end{table*}

\begin{table*}
\caption{The observed parameters and estimated masses for Type IIP
  supernovae that have mass estimates derived from
  observations of their host clusters.}
\begin{tabular}{| l | l | l | l | l | l | l | l | l |}
  \hline \hline
  SN & Metallicity  & Distance & Apparent  & Absolute  & ${\rm M_{nodust}}$ & ${\rm M_{dust}}$ \\
            & /dex                & /Mpc     & magnitude & magnitude  & ${\rm /M_\odot}$ & ${\rm /M_\odot}$ \\
  \hline
  2004am & 8.7 &  3.7$\pm$0.3  & n/a & n/a & $12^{+7}_{-3}$ & n/a \\
  2004dj & 8.4 &  3.3$\pm$0.3  & n/a  & n/a  & $15^{+3}_{-3}$ & n/a \\
  \hline \hline
\end{tabular}
\label{tab:baddetection}
\end{table*}

\subsection{The maximum likelihood limits}

The masses for the detections and non--detections have been drawn from
a distribution of Type IIP progenitors with unknown parameters. If we
assume that the progenitors are drawn from a population described by
the initial mass function (IMF), the nature of the
IMF and the range of masses which explode as Type IIP SNe are the important parameters. Following
the method of \citet{Smartt2009}, we used maximum likelihood theory to
find parameters that gave the greatest probability of generating
the data. If ${P_i}$ is the probability of the $i$th detection or
non--detection being made, the likelihood ${\rm \mathcal{L}}$ is the
probability of observing the whole dataset.

\begin{equation}
{\mathcal{L}  = \prod_{i=1}^{i=N} P_i(m)}.
\end{equation}

Taking the natural logarithm converts the product into a sum and
simplifies matters because maximising ${\rm \log_e \mathcal{L}}$ is equivalent
to maximising ${\rm \mathcal{L}}$.

\begin{equation}
{\rm \log_e \mathcal{L} = \sum_{i=1}^{i=N} \log_e[P_i(m)]}.
\end{equation}

For the non--detections, the probability of the event is the
probability that a randomly chosen star has a mass between the
lower limit and the detection limit. Thus we integrate the IMF
between these limits and normalise. The IMF is generally assumed to
be describable by a power law for supersolar masses. Here $\gamma$
is the index such that the the default Salpeter law gives
$\gamma=-1.35$. The parameters to be varied are ${m_{min}}$, the
lower mass mass limit and ${m_{max}}$, the upper mass limit.

\begin{equation}
{\rm P_i \propto \int_{m_{min}}^{m_{i}} m^{\gamma-1} dm}.
\end{equation}

\begin{equation}
{\rm P_i = \frac{m_{min}^{\gamma}-m_{i}^{\gamma}}{m_{min}^{\gamma}-m_{max}^{\gamma}}}.
\end{equation}

The detection limits ${m_i}$ are the 84 per cent confidence
limits. The probability of non detection if ${m_i}$ exceeds ${m_{max}}$ is 1 and the probability if ${m_i}$ is less than ${m_{min}}$ is 0.16.

For the detections, the probability of the event is the probability
that a star has this deduced mass subject to the errors. The
distribution of the errors is unclear and the simplest way of
accounting for the uncertainty is to integrate the IMF between the
upper and lower limits set by the errors. However, this skews the
distribution towards lower masses. We follow \citet{Smartt2009} by
instead integrating the IMF from the upper limit to the predicted mass
and then in a straight line from the value of the IMF at the predicted
mass to zero at the lower limit. If the upper error mass ${m_{i+}}$
exceeds ${m_{max}}$ then the integral is truncated at
${m_{max}}$. Similarly if the lower error mass ${m_{i-}}$ is less than
${m_{min}}$ then the integral is truncated at ${m_{min}}$.

 The maximum likelihood values are of little interest without some
 measure of how probable alternatives are. To do this the confidence
 regions must be determined. For two parameters we have the 68, 90
 and 95 per cent confidence regions when $\chi=2.3,4.6,6.2$
 \citep{Press}.

\begin{equation}
{\rm \ln \mathcal{L}_{max} - ln \mathcal{L} = \frac{1}{2}\chi}.  
\end{equation}

We calculated the maximum likelihood contours when both the upper and
lower limits are varied. Initially we used only the detections. The
non--detections favour arbitrarily low values for the lower limit
because the slope of the IMF makes low--mass stars more
probable. Then, with the lower limit fixed from the detections, the
non--detections were included to see what effect they had on the upper
limit.

\begin{figure*}
\centering
\includegraphics[width=0.45\textwidth]{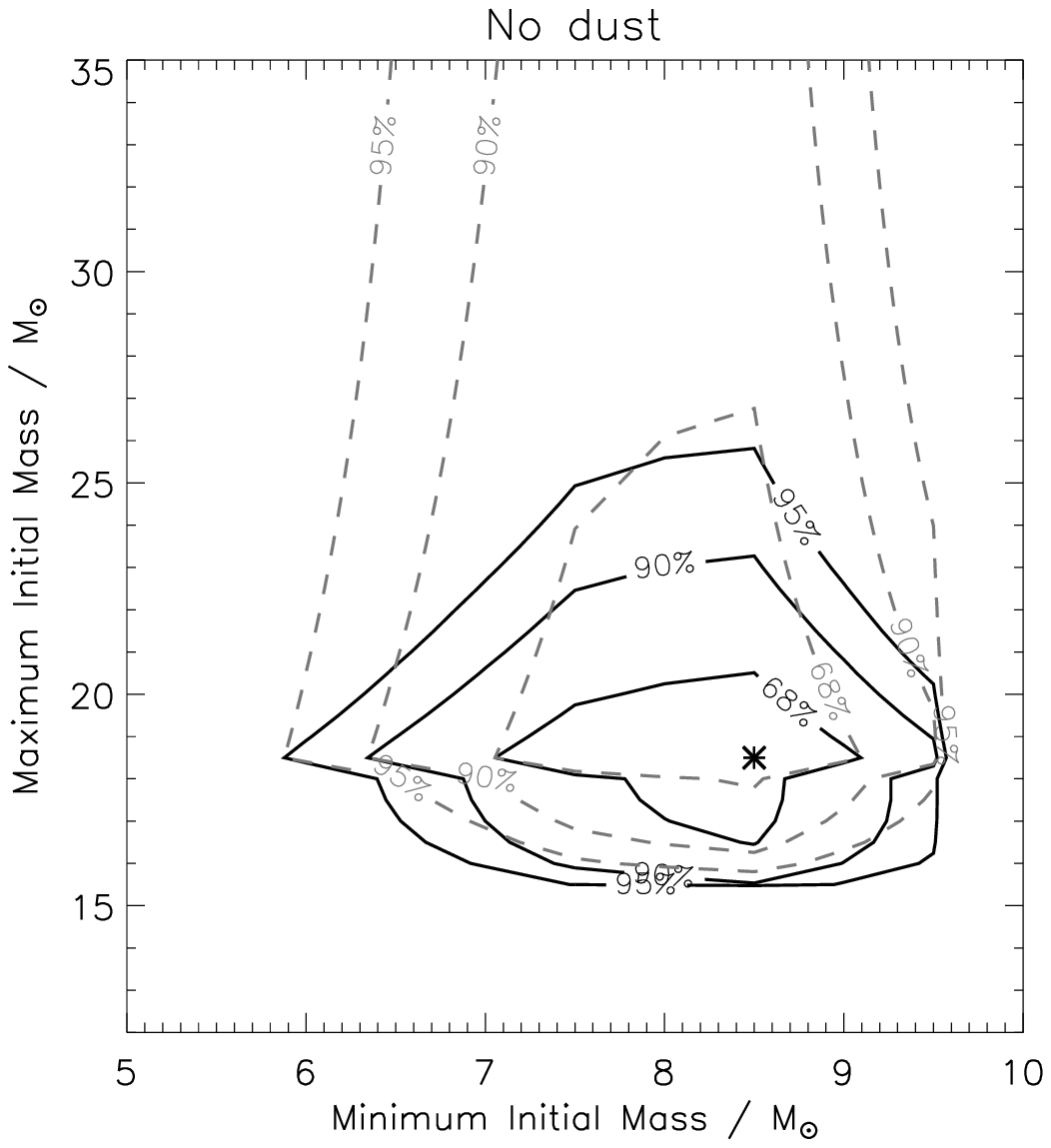}
\includegraphics[width=0.45\textwidth]{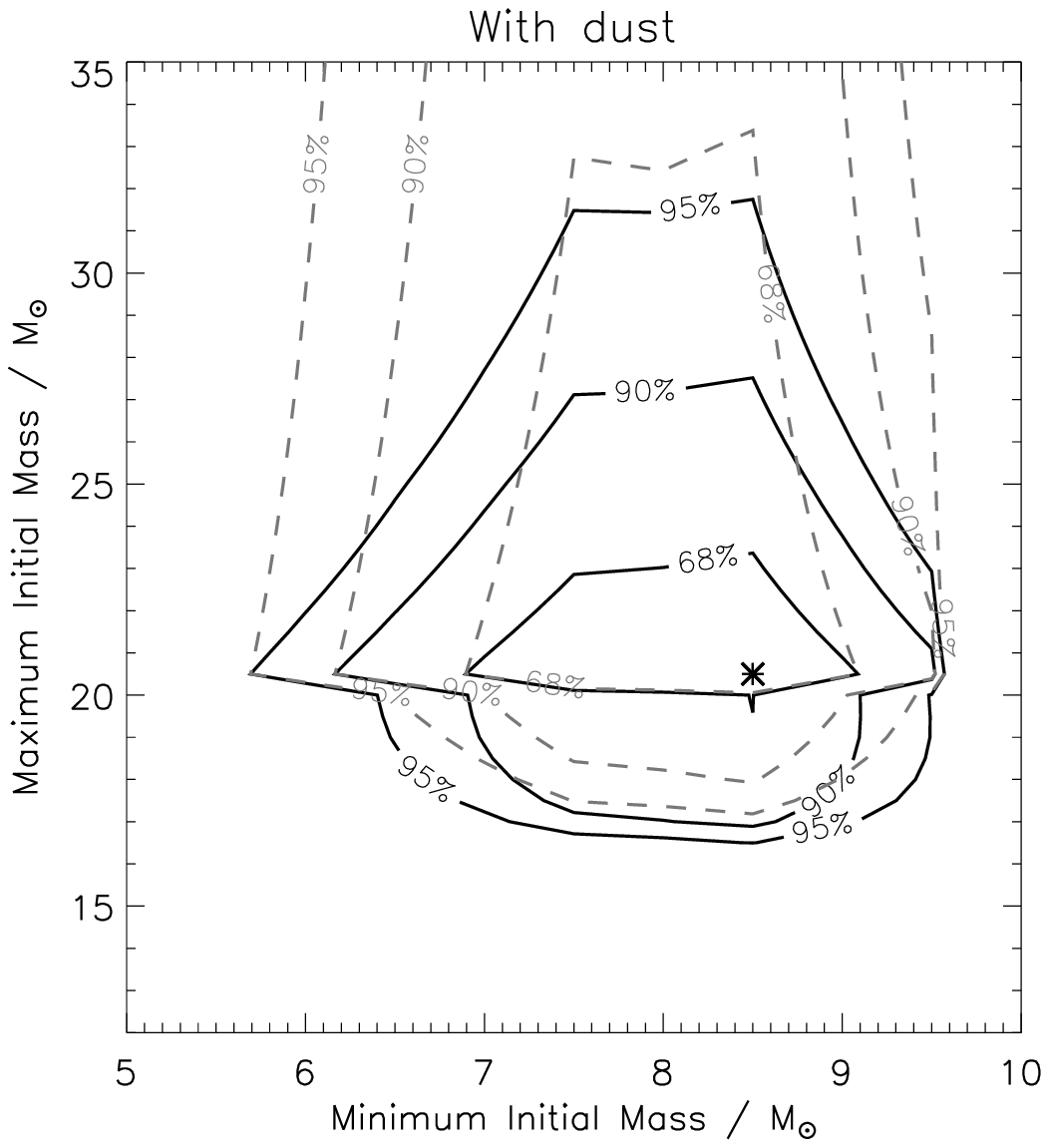}
\caption{Maximum likelihood contours. The dashed lines are for the
detections only and the solid are when the non--detections are
included.}
\label{fig:dust12}
\end{figure*}

The contrast between the two models is shown in
Figure~\ref{fig:dust12}. The models without dust predicted an upper
limit of ${\rm 18^{+2}_{-2} M_{\odot}}$ and a 95 per cent confidence
limit of 25 ${\rm M_{\odot}}$. However, the dust models give ${\rm
  21^{+2}_{-1} M_{\odot}}$ and, more significantly, a 95 per cent
limit of more than 30 ${\rm M_{\odot}}$. This means that we can no
longer say with certainty that there is population of red supergiants
that do not end as Type IIP SNe. This upper limit is
  consistent with that obtained by modelling the
  population of SN progenitors and accounting for binary evolution, i.e. \citet{2011MNRAS.412.1522S} and \citet{2011MNRAS.tmp..692E}.

\begin{figure*}
\centering
\includegraphics[width=0.45\textwidth]{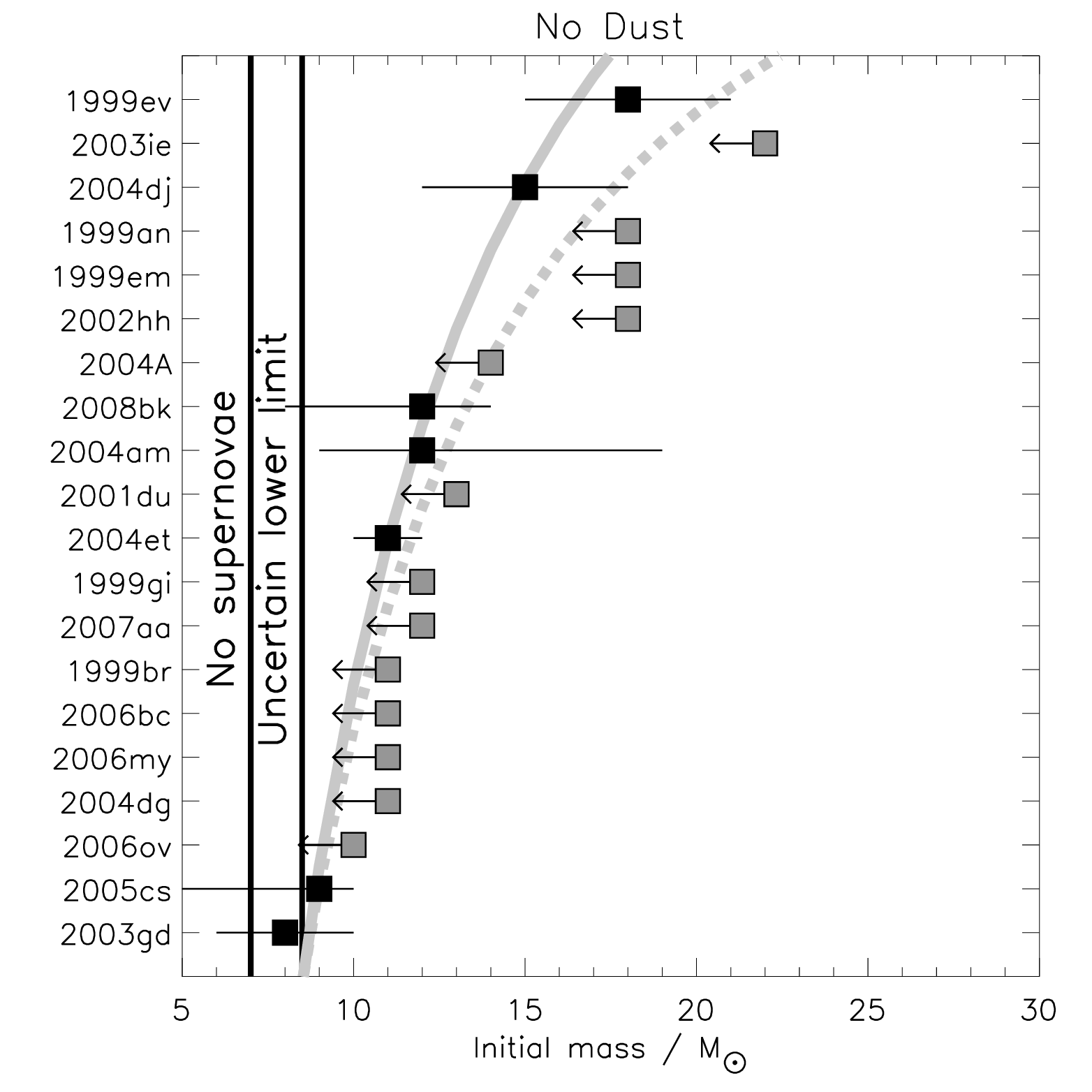}
\includegraphics[width=0.45\textwidth]{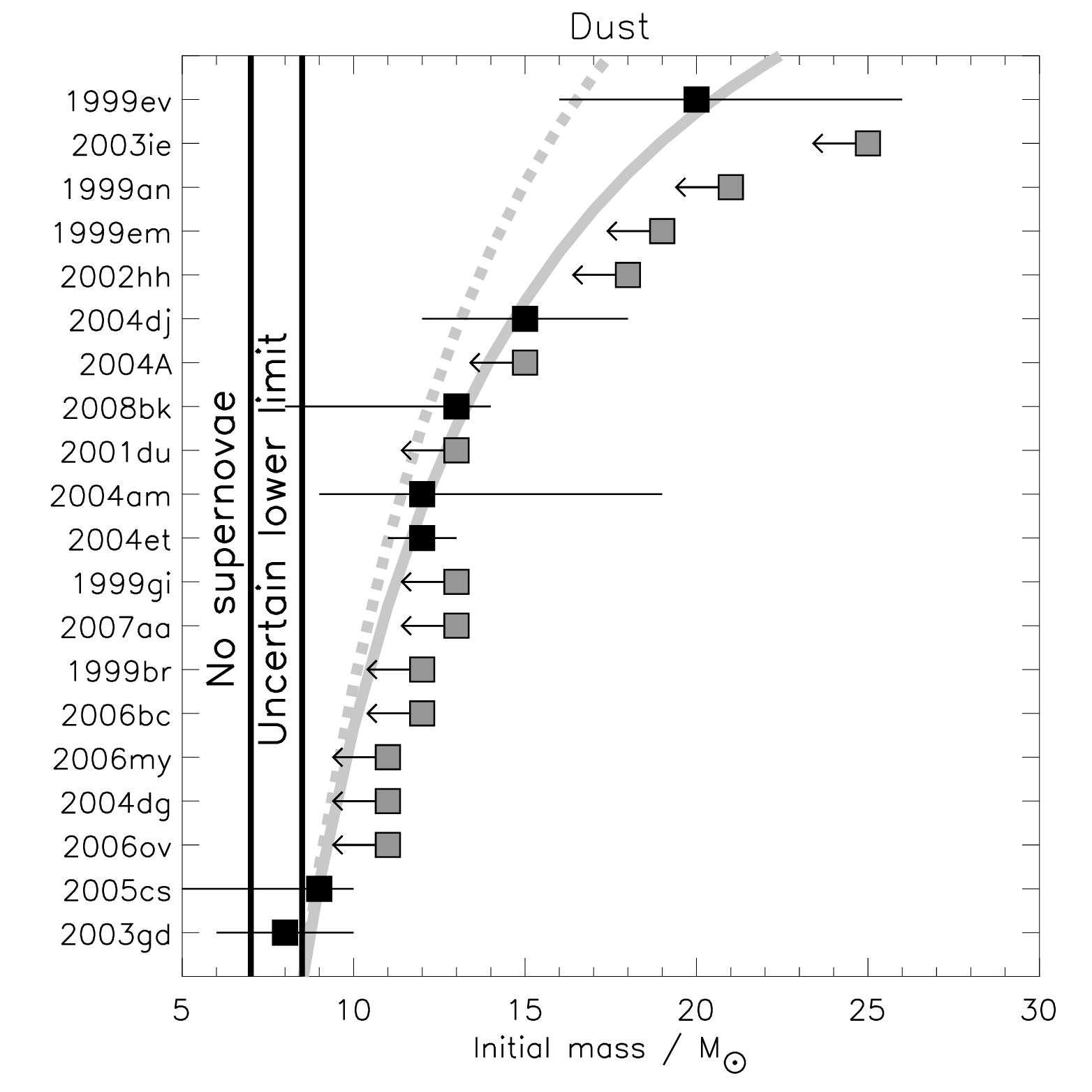}
\caption{Diagram showing the derived masses of progenitors compared to the expected mass distribution if the maximum Type
  IIP SNe progenitor mass is 16.5 ${\rm M_\odot}$ or 25 ${\rm M_\odot}$.}
\label{fig:dustB}
\end{figure*}

In Figure~\ref{fig:dustB}, we present plots of the deduced initial
masses of the progenitors with and without dust, in order of
increasing mass. They are contrasted with the curves representing a
population of stars following the Salpeter distribution, with upper limits of 16.5 and 25 $M_{\odot}$. In both cases the lower limit is
8.5 $M_{\odot}$. It should be noted that SN 1999ev, the most massive
progenitor, undergoes the greatest change in predicted mass when dust
is considered, from $18^{+3}_{-3}$ to $20^{+6}_{-4}$ ${\rm
  /M_\odot}$. No other star is more influential in deducing the upper
limit and our results will be more certain when we have more similarly
massive stars.

Recently a progenitor to SN 2009kr was identified \citep{Fraser} but
it is not clear yet whether it is a single or a binary star, a Type
IIP or a Type IIL \citep{2010ApJ...714L.254E}. If it is a single star,
the observations of ${\rm M_V=-7.6\pm0.6}$ and ${\rm
  M_V-M_I=1.1\pm0.25}$ imply initial masses of ${\rm 21^{+3}_{-4}
  M_{\odot}}$ and ${\rm 23^{+4}_{-5} M_{\odot}}$ from the models
without and with dust. However, the star is more of a
yellow supergiant and may have evolved through the red supergiant
phase \citep{2010ApJ...714L.254E}, in which case our models would not
be appropriate.

\section{Discussion}

\begin{figure}
\centering
\includegraphics[width=0.5\textwidth]{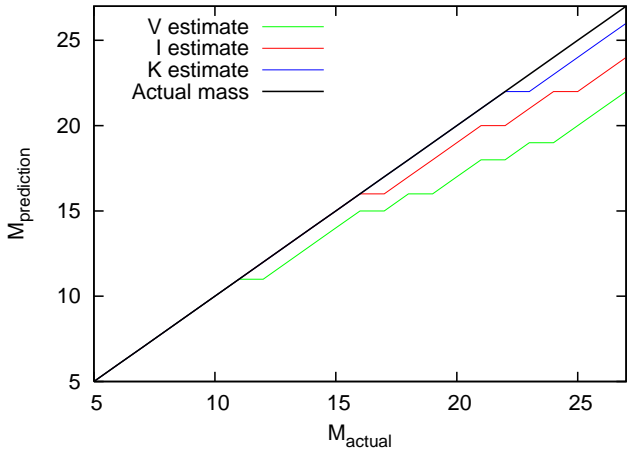}
\caption{Actual compared with deduced initial masses of red supergiants for theoretical observations in $V$, $I$, and $K$. Models are from the $Z=0.02$ series.}
\label{fig:alias}
\end{figure}

We have presented evidence that the red supergiant problem is caused
by aliasing of the higher masses. This is illustrated in
Figure~\ref{fig:alias} where we have plotted the initial mass of red
supergiants at solar metallicity against the mass that would be
deduced if we were to take the magnitudes of the dust--extincted
models and compare them with the mass--luminosity relation from the
models without dust.

The model for circumstellar dust is fairly crude, based on a best fit
between luminosity and dust production from which there is
considerable deviation. It generates an average extinction of never
more than about ${1 \, \rm A_V}$, whereas observations have revealed
stars with several times that value \citep{2005ApJ...634.1286M}.

Detailed models of dusty winds have shown that they do not maintain
spherical symmetry but form transient cap structures
\citep{Woitke}. This implies that that the amount of extinction varies
with the line of sight and over time. This variability was observed by
\citet{Massey2009}, who found an average change of 0.5 mag in the $V$
band in a sample of red supergiants in M31. This change occurred after
only three years. Significantly, there was no change in the $K$ band,
strongly implying that the change in the magnitudes was driven by
variable extinction. It is also likely that dust production varies
with metallicity but we have not taken this into account.

The main achievement of our dust model is to show that, even with an
unnaturally unvarying but realistic amount of dust production, the red
supergiant problem ceases to be. The increased extinction of the
higher--mass models introduces such uncertainty into the observations
as to make it impossible to confidently set an upper mass limit lower
than 25--30${\rm\, M_\odot}$, the red supergiant upper limit.

The upper mass limit could be determined more precisely by obtaining
more pre--explosion images in the infrared, where the effect of
extinction is much less. Alternatively, good spectroscopy of the
detected progenitors could be used to calculate the total
extinction. This would require waiting for one of the relatively small
number of nearby well--studied red supergiants to explode. Models of
the SNe themselves are also likely to yield more progenitor masses in
the future and recent work in that direction has been encouraging
\citep{2011MNRAS.410.1739D}. Until then the uncertain but significant
circumstellar extinction means that we do not need to look for
alternatives to Type IIP SNe for the death of red supergiants.

\section*{Acknowledgements}
JJW and JJE would like to thank the anonymous referee for their
suggestions. They have led to a much improved paper. JJW thanks the
STFC for his studentship and JJE is supported by the IoA's STFC theory
rolling grant. The authors would also like to thank Stephen Smartt and
Christopher Tout for discussion and comments.

\bibliographystyle{mn2e}
\bibliography{star}

\end{document}